\documentstyle[aps,preprint]{revtex}
 
\tightenlines 
\begin{document}

\title{Generic slow relaxation in a stochastic  
sandpile} 
     
\author{Ronald Dickman$^{*}$
}
\address{ 
Departamento de F\'{\i}sica, ICEx, 
Universidade Federal de Minas Gerais, 
Caixa Postal 702, 
30161-970 Belo Horizonte, Minas Gerais, Brazil
\\ 
} 
\date{\today} 
 
\maketitle 
\begin{abstract} 
Simulations of a stochastic fixed-energy sandpile
in one and two dimensions 
reveal slow relaxation of the order parameter, 
even far from the critical point.  
The decay of
the activity is best described by a
stretched-exponential form.
The persistence probability (for a
site not to have toppled up to time $t$), also exhibits
stretched-exponential relaxation.  The results
suggest a connection between sandpile models and 
structural glasses.
\vspace{2em}

\vspace{2em}

\noindent Email address: dickman@fisica.ufmg.br
\end{abstract} 
 
\date{\today} 
 
\newpage
Sandpile models have attracted much interest in recent years,
as paradigms of scale-invariance in the apparent 
absence of tuning parameters \cite{btw,ggrin,dhar99,bjp},   
and as intriguing examples of absorbing-state phase transitions
\cite{AIP,tb88,vz,carlson,fes2d,manna1d,lubeck}.   
Most studies of sandpiles have focused 
on the scale-invariant stationary state under
slow driving \cite{dhar99,priezzhev}, or on
scaling properties in the vicinity of the 
absorbing-state transition 
\cite{rossi,priezba,mnrst,mancam,manft}.
Relatively little attention has been given to 
the dynamic properties of sandpiles away from the critical point.

A central feature of sandpile models 
is the presence of a conserved field, the density of particles.
This field couples to the activity density, which is the order parameter.  
When, as in the case of FES, the conserved field is frozen in the
absence of activity, the critical behavior is expected to fall in
a universality class  distinct from that of directed 
percolation (DP) \cite{rossi}.  (In the absence of such a conservation law,
DP is generic for absorbing-state phase transitions \cite{janssen,gr82}.  
Relaxation of the order parameter to its quasi-stationary value
in a sandpile at the {\it critical} density is best characterized as
a power-law \cite{fes2d}, but with certain anomalies in the one-dimensional
case \cite{manna1d}.  Besides having an effect on the
critical behavior, it is reasonable to expect the
conservation law to modify relaxation properties away from
the critical point as well.  

In simple models exhibiting an absorbing-state
phase transition, such as the contact process \cite{marro},
relaxation to the stationary state is expected to be monotonic and exponential,
away from the critical point \cite{liggett}.  The simulation
results reported below show
that in sandpiles, off-critical relaxation is considerably slower, following
a stretched-exponential or, in certain cases, algebriac decay.
Stretched-exponential functions have been
reported for {\it avalanche size} distributions in experiments 
on sand and rice piles \cite{jeager,frette}, 
and in granular avalanche models \cite{head}, but not, to my knowledge,
in the context of sandpile relaxation dynamics.
 
I study a variant of Manna's sandpile \cite{manna}, 
defined on a lattice of $L^d$ sites (in $d$ dimensions), 
with periodic boundaries. The 
configuration is specified by the number of particles
$z_i = 0, 1, 2,...$ at each site; sites with $z_i \geq 2$ are said to be 
{\it active}.  A Markovian dynamics is defined by the toppling rate, which is 
unity for all active sites, and zero for sites with $z_i < 2$.  
When a site $i$ topples, it sends two particles to
adjacent sites ($z_i \to z_i - 2$); the 
particles move independently to randomly  
chosen nearest neighbors $j$ and $j'$ ($j$, $j' \in \{i+1,i-1\}$ in the
one-dimensional case).   
Thus $j=j'$ with probability $1/2d$.  The dynamics conserves
the number of particles, $N$, which is fixed by the initial
configuration.

The system evolves via a sequential dynamics: 
the next site to topple is  
chosen at random from a list of active sites;
the time increment associated with  
a toppling is $\Delta t = 1/N_A$, where $N_A$ is the number of active  
sites just prior to the event. 
Initial configurations are generated by 
distributing $\zeta L^d$ particles randomly over the lattice, 
yielding an initial distribution that is spatially homogeneous 
and uncorrelated. 
Previous studies confirm the existence of a continuous
phase transition from an absorbing to an active phase at 
a particle density $\zeta_c = 0.94885$ in one dimension \cite{manna1d},  
and 0.71695 in 2-$d$ \cite{fes2d}.  For $\zeta < \zeta_c$, the
stationary value of the order parameter (the density of active
sites) is zero.
 
I begin with the simplest case, relaxation of the order
parameter $\rho$ in one dimension, for $\zeta < \zeta_c$.  
For $\zeta = 0.5$,
far below the critical value, the pattern of relaxation is
essentially the same for systems of 1000 or more sites.  Figure 1
shows that $\rho (t)$ (obtained from averages over $5 \times 10^5$ 
realizations of a system of 5000 sites),
is well described by a stretched exponential, 
$\rho(t) \sim \exp[- (t/t_0)^\beta]$ with $\beta = 0.45$ ($t_0$  
represents a characteristic timescale for relaxation).
In this and subsequent analyses, the exponent $\beta$
is determined using the criterion of zero curvature, in the asymptotic
region of the graph of $\ln [\rho(t) - \overline{\rho}] $ versus
$t^\beta$ ($\overline{\rho} $ is the asymptotic activity density,
which is of course zero for $\zeta < \zeta_c$).  A rough estimate
of the uncertainty in $\beta$ values is $\pm 0.02$.
%rh5

Next I examine the relaxation for 
$\zeta = 0.9$, much nearer the critical point.  There are now considerable
finite-size effects and much larger systems (up to $10^5$ sites) are
required to observe the asymptotic behavior.  For small systems
$\rho(t)$ appears to decay faster than a stretched exponential, but
as $L$ is increased, the long-time slow relaxation develops.  For
$L = 10^5$, $\rho(t)$ again follows a stretched
exponential, with $\beta = 0.39$.
% rh9
It should be emphasized that while the simulation data cannot
be interpreted as {\it proving} stretched-exponential relaxation,
they do serve to rule out definitively both exponential and
power-law relaxation of $\rho(t)$ in the subcritical regime.

In the supercritical regime, the relaxation 
of the order parameter to its stationary value $\overline{\rho}$
is nonmonotonic.  $\rho(t)$ decays rapidly at first,
and then slowly approaches the
stationary value from below.  For $\zeta = 1$, I find 
$\overline{\rho} = 0.118222(8)$, with 
$\Delta \rho \equiv \overline{\rho} - \rho(t)$ decaying
asymptotically as a power-law, $\sim t^{-0.54}$ (see Fig. 2). 
%(rh1.dmg)
The {\it initial} decay is again well described by a stretched
exponential, with an exponent $\beta = 0.28 $.
For somewhat higher densities ($\zeta = 1.2 $ and 1.3), the
asymptotic approach to $\overline{\rho}$ (from below), is
again via a power law, with an exponent of 0.5 - 0.52.

I also studied a variant of the stochastic sandpile introduced in 
Ref. \cite{manft}, in which the toppling rate at a site with
$z$ particles is $z(z\!-\!1)$.  (In this case $\zeta_c =0.9493(2)$.)
For $p \!=\! 1/2$, the decay to $\rho = 0$ is described with 
high precision by a stretched exponential with $\beta = 0.475(25)$.
In the supercritical regime the approach to the stationary value
is nonmonotonic, similar to that observed above.
Unlike the constant-rate model, however, the relaxation is well
described by an expression consisting of two stretched
exponentials:
\begin{equation}
\rho (t) = \overline{\rho} + A_1 \exp[-\lambda_1 t^{1/2}]
- A_2 \exp[-\lambda_2 t^{1/2}] ,
\label{se2}
\end{equation}
with $A_1$, $A_2$, $\lambda_1$, and $\lambda_2$ all positive
constants ($A_1 \gg A_2$, and $\lambda_1 \gg \lambda_2$).

Slow relaxation appears to be robust under changes in the
toppling rate. Of equal interest is the nature of relaxation
in two or more dimensions.  
At a density of
$\zeta = 0.5$ (well below $\zeta_c = 0.71695(5)$),
studies using lattices of $L^2$ sites, with
$L= 40$, 80, and 160, reveal that the order parameter decays
as a stretched exponential, but with an exponent 
$\beta = 0.81$.
Nearer the critical point ($\zeta = 0.7$), finite-size effects
are prominent, as was found in the one-dimensional case.  Studies
using system sizes $L$ of up to 1280 again confirm stretched-exponential 
decay with $\beta = 0.81$, as shown in Fig. 3.
% rh2d7.dmg  

The relaxation above the critical density (in a system of
size $L=160$ at $\zeta = 0.75$; $\overline{\rho} = 0.04995$), is
nonmonotonic, as in one dimension.
The excess density is again well described by the difference between
two stretched exponentials, as in Eq. (\ref{se2}), but in this case
the exponent $\beta$ associated with the early decay is about 0.4,
while the long-time approach to $\overline{\rho}$ (from below)
is characterized by an exponent of 0.8 (see Fig. 3, inset).  

As we have seen, relaxation of the order parameter 
in the stochastic sandpile is
characterized by stretched-exponential (or in some cases,
power-law) functions.  Recently, O'Donoghue and Bray \cite{bray01}
demonstrated stretched-exponential decay of the {\it persistence
probability}, i.e., that a given site has never been visited by a
diffusing particle, in certain one-dimensional reaction/diffusion
processes.  This suggests a study of persistence in the sandpile.
In fact, since a site that is initially below threshold for toppling
{\it must} be visited by another diffusing particle or particles before it
can topple, persistence seems particularly relevant to sandpile
relaxation dynamics.  (Studies of dynamic critical exponents for
persistence in conserved lattices gases, which are closely related
to sandpiles, were reported by L\"ubeck \cite{lubeck02}; here,
however, we focus on the dynamics away from the critical point.)

In studying persistence in a sandpile, it appears useful to group
sites according to their initial occupation number $z$. Let
$p(t;z)$ denote the persistence probability of sites whose initial
height is $z$.  For $z = 0$ or 1, I define $p(t;z)$ as the probability that
a site has not toppled up to time $t$.  For $z > 2$, however, such a
definition is not very interesting, since (in the constant-rate model),
the probability not to have toppled up to time $t$ is simply $e^{-t}$.
For $z \geq 2$, therefore, $p(t;z)$ is defined as the probability not
to have toppled a {\it second time}.  I study the one-dimensional
fixed-rate model with particle densities $\zeta = 0.5$, 0.9, and 1,
as above.

For $\zeta < \zeta_c$, all activity ceases after a finite time, so that
the persistence probability approaches a nonzero value $\overline{p}(z)$
at long times.
For $\zeta = 0.5$, the asymptotic relaxation of $p(t;z)$ to
$\overline{p}(z)$ again follows a stretched exponential; the best
estimates for the exponent $\beta$ are 0.42 for $z=0$ and $\beta = 0.39$
for $z = 1$, 2 and 3.
%               (pst5.dmg)
For $\zeta = 0.9$, a study using $L = 10^4$ yielded $\beta \simeq 0.44$,
0.47, 0.49, and 0.41 for $z = 0$, 1, 2, and 3, respectively.
%               (pst9.dmg)
Finally, for $\zeta = 1$, the asymptotic decay of the persistence 
probability (studied on a ring of $2 \times 10^4$ sites)
follows a stretched exponential with $\beta$ values of 0.42, 0.41, 
0.38, and 0.45 for the various $z$ values.  The persistence
probabilitites for $\zeta = 0.9$ and $\zeta =1$ are shown in Fig. 4.
%             (pst1.dmg)
(In all of these studies,
the exponents for $z = 2$ and 3 are less certain, due to poorer statistics, 
and to the fact that stretched-exponential behavior sets in later
than for $z=0$ or 1.  I find no clear trend in
the estimated exponent value, either with $z$ or with $\zeta$.)
The foregoing results can be summarized as indicating that 
in one dimension, the asympotic
relaxation of the persistence probability $p(t;z)$ follows a stretched
exponential with an exponent $\beta$ of about 0.42.

The simulation results indicate that the relaxation of the order parameter
and of the persistence probability, in a stochastic sandpile away from
its critical point, generically exhibits stretched-exponential scaling.
The exponent $\beta$ generally takes values in the neighborhood of
0.4 in the one-dimensional case, while in two dimensions (for which
only order-parameter relaxation was studied) the value is about 0.8.
Stretched-exponential relaxation
may be understood as a consequence of the {\it depletion}
of active elements, leading to a decreasing relaxation rate, 
$(1/\rho)|d\rho/dt|$, as time goes on.  
The larger exponent in two dimensions is plausible, given 
the larger number of paths in configuration space.  
Detailed explanations of the 
streteched-exponential form, and of specific exponent values, 
remain as open challenges.

The observation of stretched-exponential relaxation suggests a
connection with glassy dynamics.  Superficially, the
sandpile model seems to have little
connection with dynamics of a dense fluid, but the two problems
are related at a more abstract level.  In dense fluids, not all the empty space in
the system is available for particle movement, leading to a highly
cooperative dynamics \cite{poole,doliwa}.  Some larger grouping
of ``voids" is required for relaxation on scales beyond that of the local cage.
(For simplicity, I frame the analogy
in terms of a hard-sphere fluid, in which the
relevant variable is density not temperature.)
Similarly, in a sandpile not all particles are available for movement
(and relaxation): only those with companions in the same cell are mobile.
These observations suggest a parallel between {\it particles} in
the sandpile and {\it parcels of unoccupied space} in a dense fluid.
The dynamic arrest in the fluid, when the density approaches an
(apparent) transition value, corresponds to the absorbing-state 
phase transition in the sandpile.  
The associated order parameter - mobile free volume in the fluid -
vanishes in the absorbing phase.  
(The total unoccupied volume is of course conserved, but the
fraction that is mobile or ``active" is not.)
From this vantage, the liquid-glass transition corresponds to a 
{\it dynamic} transition, akin to directed percolation, of free volume, as opposed to
the static percolation transition suggested some time ago by Cohen and
Grest \cite{cohengrest}.
While the sandpile models
studied to date seem too simplistic to capture the dynamics of
a dense fluid near the glassy state, the results of the present study
suggest that it would be worthwhile pursuing this analogy by developing
somewhat more realistic particle models, and studying diffusion and the
response to external driving, in the hope of gaining further insight
into real glasses.
 \vspace{1em}

\noindent{\bf Acknowledgments}
\vspace{.5em}

\noindent I thank Kim Christensen and Francisco F. Araujo Jr. for helpful discussions.   
This work was supported by
CNPq, CAPES, and FAPEMIG, Brazil.

\newpage 
\noindent FIGURE CAPTIONS 
\vspace{1em} 
 
\noindent FIG. 1.  Relaxation of the order parameter in the one-dimensional
sandpile, $\zeta \!=\! 0.5$, $L \!=\! 5000$.  Inset: semi-log plot of the same data.
\vspace{1em} 
 
\noindent FIG. 2.  Relaxation of the order parameter in the one-dimensional
sandpile, $\zeta \!=\! 1$, $L \!=\! 2 \times 10^4$.  Inset: log-log plot of
the excess $\Delta \rho = \rho(t) - \overline{\rho}$.
\vspace{1em} 

\noindent FIG. 3.  Relaxation of the order parameter in the two-dimensional
sandpile, $\zeta \!=\! 0.7$, $L= 1280$.  Inset: $\ln |\Delta \rho| $ versus
$\ln t$ for $\zeta = 0.75$; points: simulation; solid curve: fit using 
two stretched exponentials as described in text.
\vspace{1em} 

\noindent FIG. 4.  Persistence probabilities $p(t;z)$ for $z = 0$, 1, 2, and 3
(upper to lower) in the one-dimensional model with $\zeta = 1$.  Inset: a
similar plot for $\zeta = 0.9$.
\vspace{1em} 
 
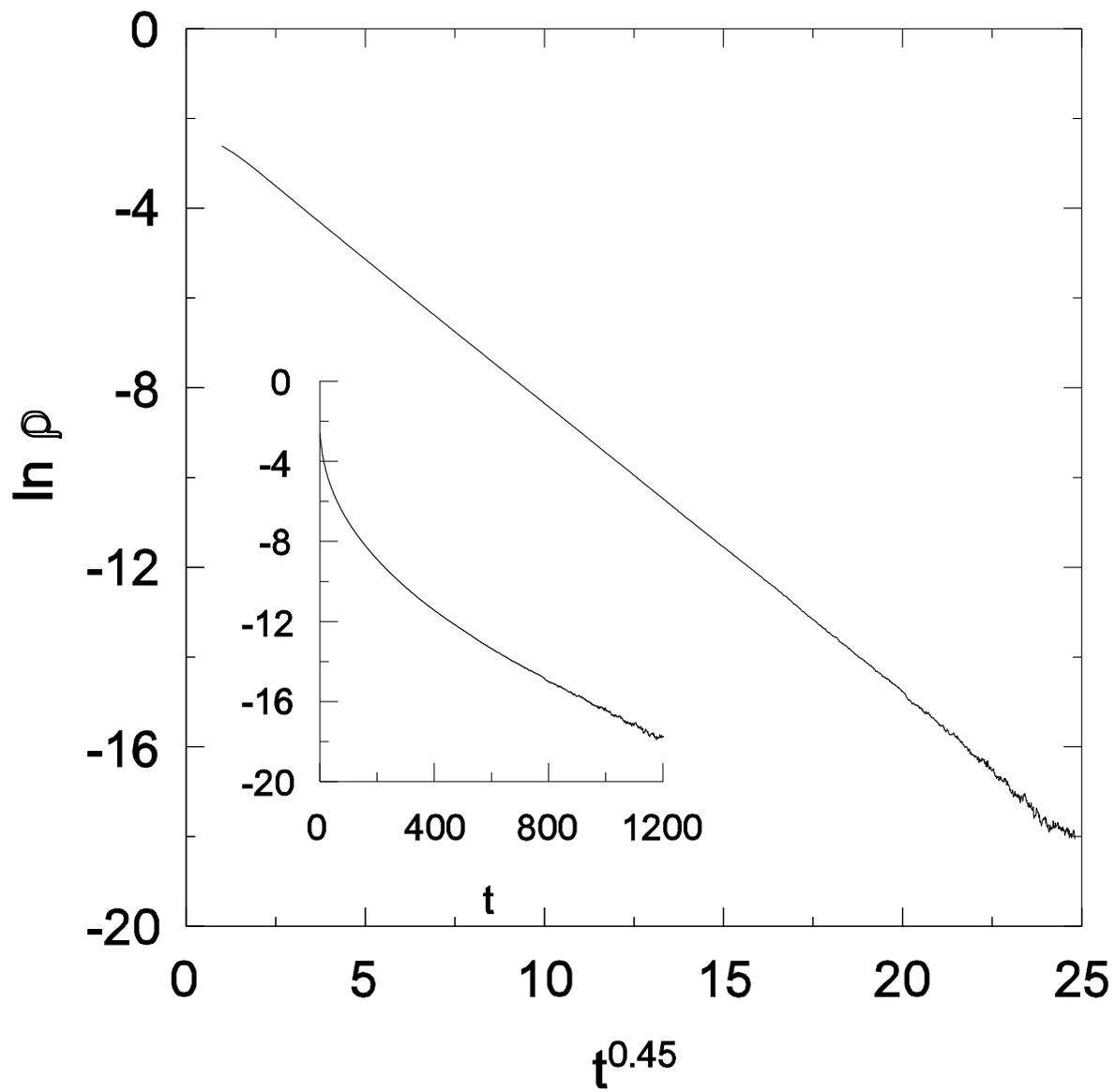
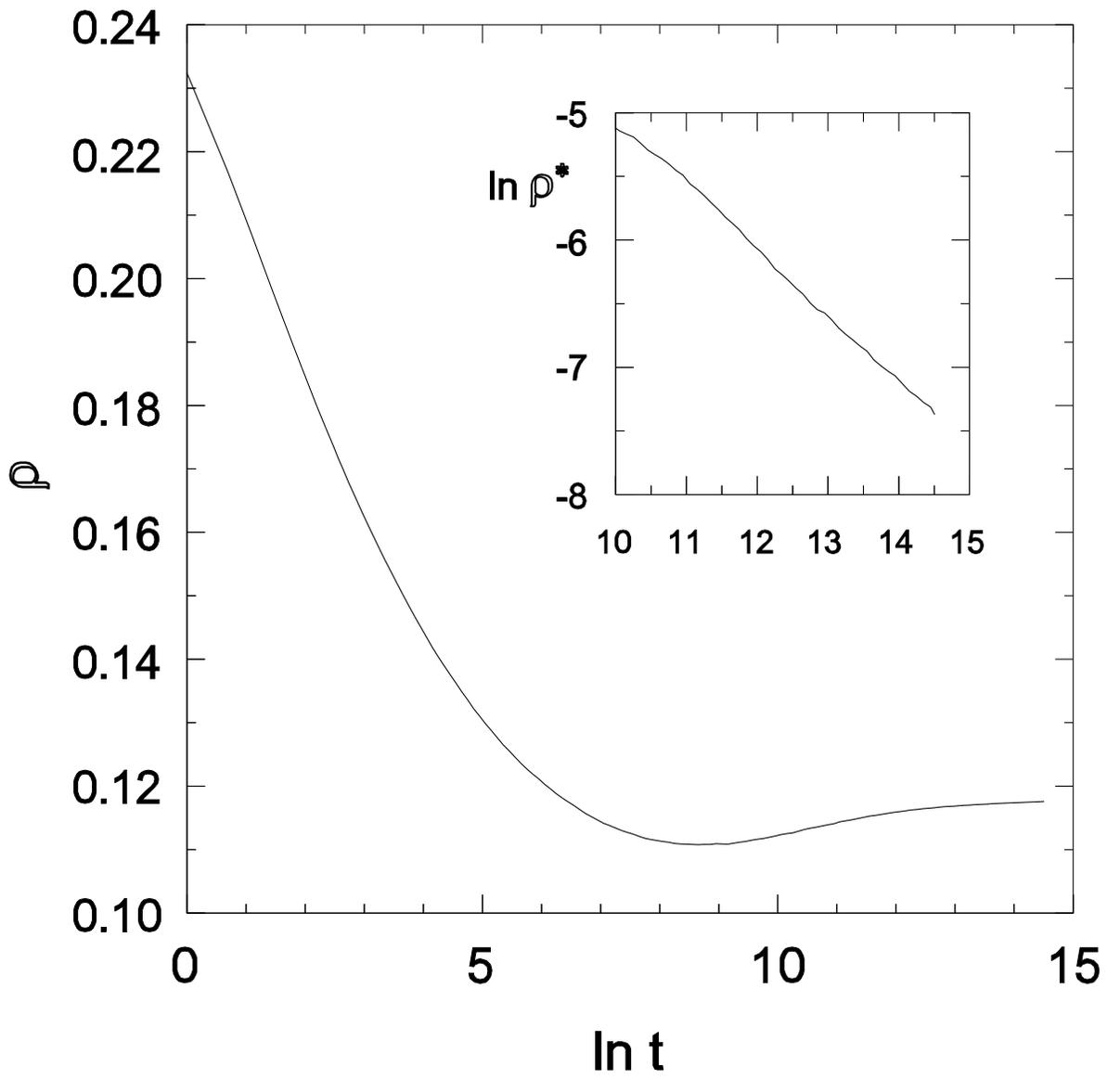
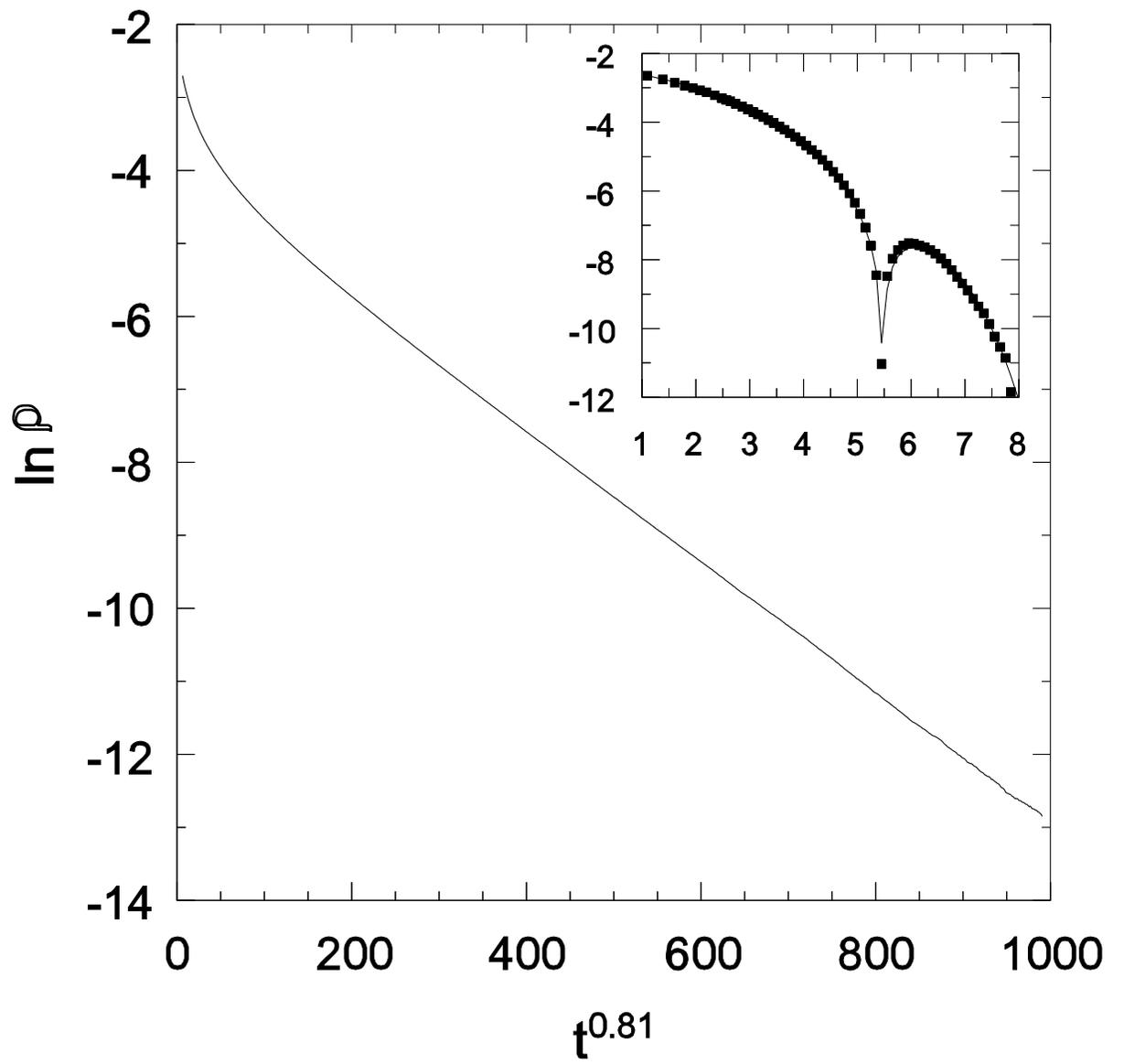
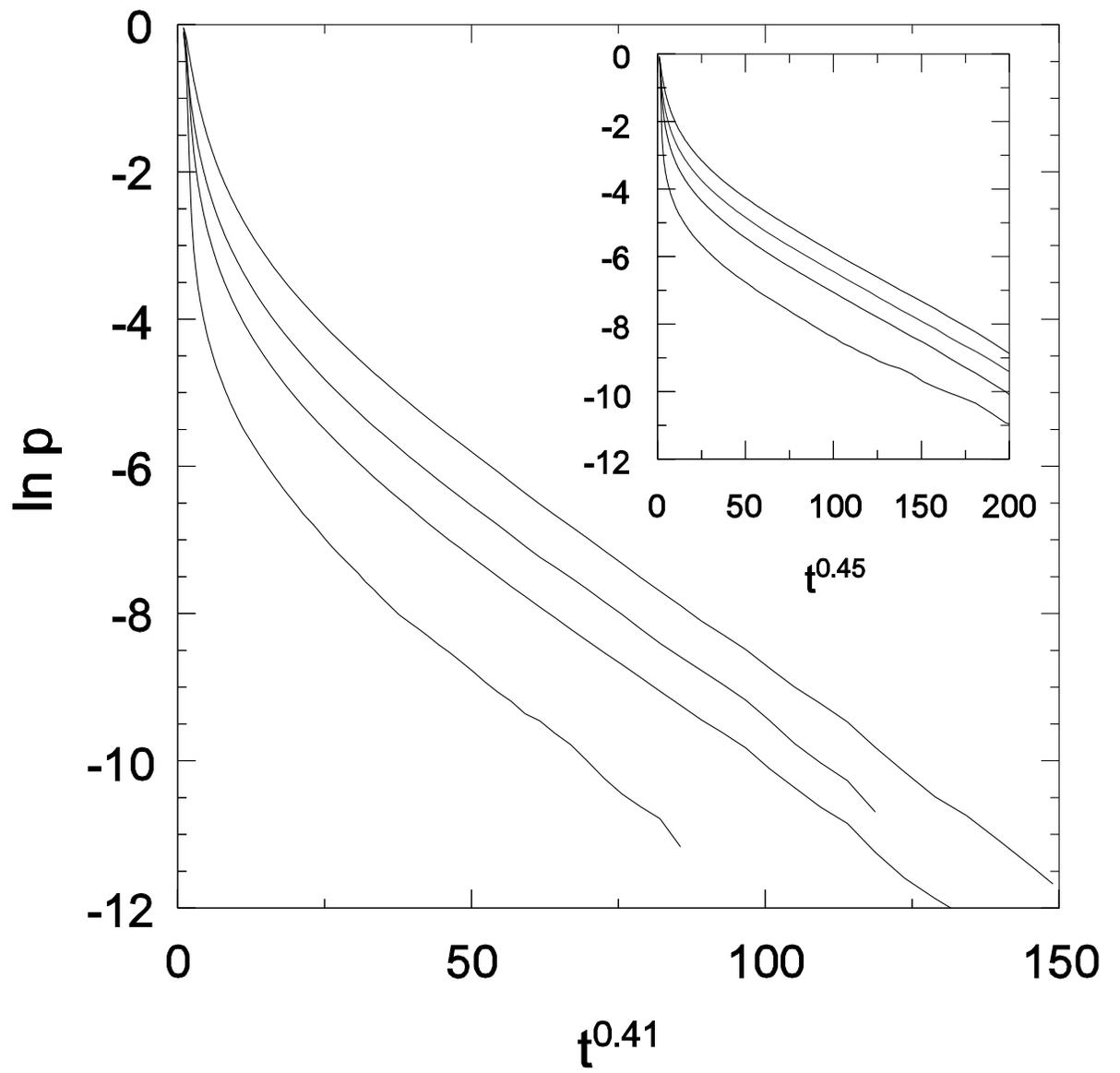
\end{document}